\documentclass[prd,showpacs,preprintnumbers,amsmath,amssymb]{revtex4}% Physical Review B

\usepackage{graphicx}% Include figure files
\usepackage{dcolumn}% Align table columns on decimal point
\usepackage{bm}% bold math

\newcommand{\nn}{\nonumber}
\newcommand{\bi}{\bibitem}
\newcommand{\be}{\begin{eqnarray}}
\newcommand{\ee}{\end{eqnarray}}

\catcode`\@=11
\def\lsim{\mathrel{\mathpalette\@versim<}}
\def\gsim{\mathrel{\mathpalette\@versim>}}
\def\@versim#1#2{\vcenter{\offinterlineskip
\ialign{$\m@th#1\hfil##\hfil$\crcr#2\crcr\sim\crcr } }}
\catcode`\@=12

\begin{document}

%\tightenlines

\draft
\preprint{KANAZAWA-03-32}
\preprint{MPP-2003-77}
\title{Softening the Supersymmetric Flavor Problem\\
 in Orbifold GUTs}
\author{Yuji Kajiyama$^1$}
\author{Jisuke Kubo$^{1,2}$}
\author{Haruhiko Terao$^1$}

\affiliation{$^1$Institute for Theoretical Physics, Kanazawa
University,
  Kanazawa 920-1192, Japan\\
$^2$Max-Planck-Institut f\"ur Physik,
  Werner-Heisenberg-Institut\\
D-80805 Munich, Germany}

%\date{\today}

\begin{abstract}
The infra-red attractive force of the bulk gauge interactions
is applied to soften the supersymmetric flavor problem
in the orbifold SU(5) GUT of Kawamura. 
Then this force aligns in the infra-red regime the
soft supersymmetry breaking terms out of their anarchical disorder at
a fundamental scale, in such a way that flavor-changing neutral
currents as well as dangerous CP-violating phases are 
suppressed at low energies. It is found that this dynamical 
alignment is sufficiently good compared with the current 
experimental bounds, as long as the diagonalization matrices
of the Yukawa couplings are CKM-like.
\end{abstract}


%\narrowtext
%\twocolumn

\maketitle

The major success of grand unified theories (GUTs) based
on softly broken $N=1$ supersymmetry (SUSY) is
the gauge coupling unification that makes possible
to predict  one of the three gauge couplings
of the standard model (SM) \cite{susy}.
Despite the success, there are still problems that
we are faced with and deserve theoretical attentions.
One of them is the problem of the ``doublet-triplet splitting'':
If the SM Higgs doublet is embedded to a larger representation
of a GUT, there will be colored partners of the SM Higgs doublet.
These colored partners will cause the
fast nucleon decay in general, unless there are extremely heavy
\cite{d5,murayama}.
So, the question how one can arrange
without fine tuning of parameters
to keep the Higgs doublet light while making the colored Higgs
superheavy
should be answered.
 Recently, Kawamura \cite{kawamura1}
 suggested a simple idea in five  dimensions
 that is compactified on
$ S^1/(Z_2\times Z_2^{\prime})$ \cite{barbieri}.
 He showed that the zero modes of the
 $N=2$  gauge supermultiplet of $SU(5)$ and two pairs of the
 Higgs hypermultiplets in ${\bf 5}$ and ${\bf \bar{5}}$
 can be so projected out
that only those that correspond to the minimal supersymmetric
 standard model (MSSM) remain as zero modes \footnote{See also
\cite{altarelli} and \cite{hall1}.}.
 So, the doublet-triplet splitting problem is shifted to that of the
 space-time geometry, which may be answered
 in a more fundamental theory that contains
 gravity \cite{witten}.
 
Another very difficult problem
 is the SUSY flavor problem:
 In its phenomenological applications,
 SUSY is introduced to protect
the Higgs mass from the quadratic divergence.
Therefore, 
the effects of supersymmetry breaking should appear
at low-energies
as soft supersymmetry breaking (SSB) terms \cite{susy}.
However, if only renormalizability is used to guide
the SSB parameters, it is possible to
introduce more than 100 new parameters into the
MSSM \cite{dimopoulos1}.
The problem is not only this large number of
the independent parameters, but also the fact that one has to
highly fine tune these parameters so that they do not cause
problems with experimental observations
on the flavor changing neutral current (FCNC) processes and
CP-violation 
phenomena \cite{fcnc-mueg,fcnc-k,fcnc-edm,fcnc-bsg,fcnc}.
There are several approaches
\cite{gauge,anomaly,gaugino,susy} to overcome this problem.
Their common feature is the assumption
 that there exists a hidden sector in which
 SUSY is broken by some flavor blind mechanism,
 and that  SUSY breaking is mediated by  flavor blind interactions
 to the MSSM sector
\footnote{In \cite{hall2,hamaguchi,babu1,kubo6}, permutation
symmetries
have been used to soften the SUSY flavor problem.}.
Another type of idea to overcome the SUSY flavor problem is to
use the infrared attractive force of
the gauge interactions
\cite{ross, karch,ns,knt,ls,abel}.
Along this line of thought, it was recently suggested in
\cite{kubo6} (see also \cite{choi1})
to introduce extra dimensions in SUSY GUTs to amplify
the infrared attractive force of gauge interactions.
It was found that this force can align the
SSB terms out of their anarchical disorder at a
fundamental scale, even if the ratio of the fundamental scale
$M_{\rm PL}$ to the GUT scale $M_{\rm GUT}$ is small
 $\lsim O(10^3)$.
 
The reason of this dynamical alignment of the SSB
parameters is simple.
The couplings in the Kaluza-Klein (K-K) theories
show power-law running behavior. Therefore the running
gauge coupling and the corresponding gaugino mass are highly
enhanced towards infra-red in asymptotically free theories.
Then the radiative corrections by the gauge interaction,
which dominate over the tree level values, make the effective
soft parameters aligned to the flavor universal forms
at low-energy.
The main assumption in \cite{kubo6} was that only the gauge
supermultiplet propagates in the bulk of the extra dimensions
to suppress the flavor dependent contributions of the Yukawa
couplings
to the RG running of the SSB parameters;
the Yukawa couplings obey only the logarithmic law of running
in this assumption.
Therefore, the assumption of \cite{kubo6} as it stands
does not fit to the orbifold GUT of \cite{kawamura1}.

In this paper, we are motivated by the
desire to combine  the  mechanism of \cite{kubo6} to solve the
SUSY flavor problem with
the  idea of \cite{kawamura1} to overcome
the doublet-triplet splitting problem.
Since in the orbifold GUT of  \cite{kawamura1}
it is essential that the Higgs hypermultiplets
also propagate in the bulk,
the Yukawa  couplings obey the power law of
running \cite{veneziano,dienes1,kobayashi1,ejiri} and nontrivially
contribute
to the RG running of the SSB parameters,
and hence can introduce
a flavor dependence in the SSB parameters.
{\footnote{Non-universalities of the soft terms induced by the Yukawa 
couplings have been studied in the four-dimensional GUT models \cite{bhs}.}}
Since, however, the Yukawa couplings of the first two
generations may be assumed to be small,
the flavor-blind infrared attractive force of
the gauge interactions are still dominant in
the running of the SSB parameters of the first two generations.
In contrast to them,  the running of the
SSB parameters of the third generation will be modified,
because the Yukawa couplings of the third generation can
be comparable with the gauge coupling in magnitude.
We therefore expect a certain
splitting between the soft scalar masses of the first two and third
generations. 
We will see that this splitting, especially by the top quark Yukawa coupling,
is sizable, however the flavor mixing masses at low energy can be consistent
with the observations on the FCNC processes.

\section{Gauge Coupling Unification in
Orbifold GUTs}
In ordinary GUTs, the gauge coupling unification
is a consequence of the unification
of the SM gauge groups into a simple unified gauge
group ${\cal G}$. In orbifold GUTs, however,
${\cal G}$  is
explicitly broken by   the boundary
condition. Therefore, the gauge coupling unification
is not an automatic consequence of  ${\cal G}$.
Below we would like to make
a quantitative consideration on the consequence of this breaking.

Let us start by  assuming that the $\beta$ functions of the SM gauge
couplings
$g_1,g_2$ and $g_3$ above the
compactification scale $\Lambda_C=1/R$ can
be written in the one-loop level
as \cite{veneziano,dienes1,kobayashi1,ejiri,hall1}
\be
\beta_i &=& \Lambda \frac{d g_i}{d \Lambda}
=
\frac{g_i^3}{16\pi^2}
\left[-b X_\delta (R \Lambda)^\delta+\Delta b_i \right],
\label{beta1}
\ee
where $X_\delta$ expresses the regularization dependent
coefficient \cite{kubo1}.
The first term represents contributions of the bulk fields,
which are common for all gauge couplings.
The second term $\Delta b_i$
results from the massless modes,
where the structure of
the massless modes depends on
the boundary condition of an orbifold model.
Eq. (\ref{beta1}) can be easily integrated, and we find
\be
\frac{1}{g_i^2}(\Lambda)&=&
\frac{1}{g_i^2}(\Lambda_0)
+\frac{1}{8\pi^2}\left\{ 
bX_\delta  [(R\Lambda)^\delta -(R\Lambda_0)^\delta]
-\Delta b_i\ln \frac{\Lambda}{\Lambda_0} \right\}.
\label{evolution1}
\ee
We  emphasize that
the unification of the SM gauge symmetry takes place
not in
$D=4$  rather in $D=4+\delta$ dimensions in which the original theory
is formulated. 
Note also that the gauge couplings
in Eq. (\ref{beta1}) are
appropriately normalized  for  four dimensions. We therefore consider
the (dimensionless) couplings $\hat{g}_i$ which
are the true expansion parameters in $4+\delta$
dimensions:
\be
\hat{g}_i &=& g_i (R\Lambda)^{\delta/2},
\label{5d-g}
\ee
where their $\beta$ functions are given by
\be
\hat{\beta}_i &=& \Lambda \frac{d \hat{g}_i}{d \Lambda}
=\frac{\delta}{2}\hat{g}_i+
\frac{\hat{g}_i^3}{16\pi^2}
\left[-b X_\delta +\frac{\Delta b_i}{(R \Lambda)^\delta}\right].
\label{5d-evolution}
\ee
So, in terms of the $4+\delta$ dimensional gauge couplings 
$\hat{g}_i$, we
see explicitly that the breaking term
$\Delta b_i$ is suppressed by
the inverse power of $\Lambda$ and
hence the unified group ${\cal G}$
recovers as $\Lambda$ goes to $\infty$.
Further, the analog of  Eq. (\ref{evolution1}) becomes
\be
\frac{1}{\hat{g}_i^2}(\Lambda)&=&
\left(\frac{\Lambda_0}{\Lambda}\right)^{\delta}
\frac{1}{\hat{g}_i^2}(\Lambda_0)
+\frac{1}{8\pi^2}
\left\{ \frac{bX_{\delta}}{\delta}
\left[ 1-\left( \frac{\Lambda_0}{\Lambda}\right)^\delta \right]
-\frac{\Delta b_i}{(R\Lambda)^\delta}
\ln \frac{\Lambda}{\Lambda_0}\right\},
\label{evolution2}
\ee
so that the difference of two gauge couplings at $\Lambda$ is
\be
\frac{1}{\hat{g}_i^2}(\Lambda)-
\frac{1}{\hat{g}_j^2}(\Lambda) &=&
\left(\frac{\Lambda_0}{\Lambda}\right)^{\delta}
\left[ \frac{1}{\hat{g}_i^2}(\Lambda_0)-
\frac{1}{\hat{g}_j^2}(\Lambda_0)\right]
-\frac{1}{8\pi^2}\frac{(\Delta b_i-\Delta b_j)}{(R\Lambda)^\delta}
\ln \frac{\Lambda}{\Lambda_0}.
\label{diff}
\ee
%Eq. (\ref{diff}) implies that
%$\hat{g}_i^2(\Lambda_0)=\hat{g}_j^2(\Lambda_0)$
%should be satisfied so that two gauge couplings
%flow into one as $\Lambda \to \infty$.
The point is that Eq. (\ref{diff}) does not imply that all three
couplings
have to coincide with each other at a single scale $\Lambda_0$
in order for the unified symmetry to recover at $\Lambda=\infty$.
This consequence seems to be rather natural, since the boundary
effect breaking ${\cal G}$ should not influence to a much shorter
length scale than the radius of the compactified dimensions.
We may call this ``asymptotic unification".

Here it may be wondered if the fundamental scale cannot
be taken much higher than the compactification scale,
since the $4+\delta$ dimensional gauge coupling seems
to exceed its strong-coupling value with
which the loop expansion becomes meaningless.
This naive dimensional observation \cite{gaugino}
follows from Eq.~(\ref{5d-g}) with keeping the four dimensional
gauge coupling to be a constant. However the
running behavior of the coupling should be taken into
account. It is indicated by Eq.~(\ref{evolution2}) that
the $4+\delta$ dimensional gauge coupling approaches
to a UV fixed point \cite{ejiri, dienes2}.
Therefore we assume that the fundamental scale can be taken
up to the Planck scale and the 1-loop RG analysis is
at least qualitatively valid there.

 On the other hand
the interpretation of Hall and Nomura
\cite{hall1} may be called ``rigid unification".
In their scheme all the gauge couplings should coincide
with each other at a single scale
and above that scale the orbifold model goes over
to a fundamental theory so that
the evolution of the gauge couplings described
by Eq. (\ref{beta1}) can be used only below that scale.
This may be a possible scenario,  but  not the  one
that is forced by the orbifold model without
knowing what the fundamental theory is.
So, at least to our understanding, the question
of what the gauge coupling unification
in orbifold GUTs means is still open.
In this paper, we would like to adopt the ``asymptotic
unification", though this notion includes the ``rigid
unification".  

The difference of a K-K GUT
(in which the boundary condition does not
break ${\cal G}$) and an orbifold model
is the boundary condition, and it appears
as  the logarithmic corrections in
the evolution of the gauge couplings
Eq. (\ref{evolution2}),
which originate from $\Delta b_i$.
As we see in Eq. (\ref{evolution2}), the effect of
$\Delta b_i$ becomes smaller and smaller as $\Lambda$
increases. So,  at the zeroth order of approximation,
we may neglect $\Delta b_i$.
At this  order the orbifold model
behaves exactly the same as the K-K model:
all the couplings meet at $M_{\rm GUT}$ and
$g_i(M_{\rm GUT})=g_{\rm GUT}$.

In practice the gauge couplings do not meet at the single
scale $M_{\rm GUT}$ when the logarithmic corrections are included.
Since this should be recovered in the limit $\Delta b_i \to 0$,
we may write
\be
\frac{1}{\hat{g}_i^2}(M_{\rm GUT})
&=&\frac{1}{\hat{g}^2_{\rm GUT}}+
\sum_k A^{i}_k \Delta b_k,
\label{corrections}
\ee
where $A$'s are $O(1)$ constants.
Therefore, unification of gauge couplings should be
disturbed only by  uncontrollable $O(g^2)$ corrections.
The main point of our analysis is to see alignment of
the soft supersymmetry breaking parameters and
the effect of the logarithmic corrections to their RG
running is slight even quantitatively. Hence we adopt
the lowest order approximation, in which the logarithmic
corrections are neglected, in this paper.

Here we would like to emphasize that even in this
approximation the power-law-running Yukawa couplings
split the SSB parameters of the first two and third
generations, as is seen later on.
So the most dominant flavor dependent effects by the
Yukawa couplings can be studied in this order.
Throughout this paper, we therefore shall
work in the lowest order of approximation 
in the scheme of ``asymptotic unification".

\section{The Kawamura Model
with Soft Supersymmetry Breaking}

\subsection{Action}
Let us start by considering the $SU(5)$ orbifold model
 proposed in \cite{kawamura1,hall1}, where
to simplify the situation, we also would like to
neglect the neutrino masses
and their mixings. 
The field content is as follows.
The $N=2$ vector supermultiplet contains
an $N=1$ gauge supermultiplet $V=(V^a~,~V^{\hat{a}})$
and an $N=1$ chiral supermultiplet $ \Sigma
=(\Sigma^a~,~\Sigma^{\hat{a}})$ in the adjoint
representation, where $a$ denotes the generators of
the SM gauge group $SU(3)\times SU(2) \times U(1)$,
and $\hat{a}$ stands for the rest of
the $SU(5)$ generators.
Two MSSM Higgs superfields are
a part of two pairs of $N=2$ hypermultiplets,
$\{~H({\bf 5})~,~H^c({\bar{\bf 5}})~\}~,
~\{~\overline{H}^c({\bf 5})~,~\overline{H}({\bar{\bf 5}})~\}$,
where the MSSM Higgs doublets $H_u$ and $H_d$ are contained in
$H$ and $\overline{H}$, i.e.,
$H=(H_C({\bf 3,1})~,~H_u({\bf 1,2}))~,~\overline{H}
=(H_{\bar{C}}({\bf \bar{3},1})~,~H_d({\bf 1,2}))$.
The parity assignment of the fields under $Z_2\times Z_2'$ is the
same 
as in \cite{kawamura1}:
\be
V^a, H_u, H_d & : & (+,+)~,~
V^{\hat{a}}, H_C, H_{\bar{C}}  :  (+,-)~,\\
\Sigma^{\hat{a}}, H_C^c, H_{\bar{C}}^c & : & (-,+)~,~
\Sigma^a, H_u^c, H_d^c  :  (-,-)~,
\ee
where the mode expansions are given by \cite{hall1}
\be
\phi_{++}(x^\mu,y) &=&\sum_{n=0}
\frac{1}{\sqrt{2^{\delta_{n,0}}\pi R}}\phi_{++}^{(2n)}\cos \frac{2n
y}{R}~,\label{phi++}\\
\phi_{+-}(x^\mu,y) &=&\sum_{n=0}
\frac{1}{\sqrt{\pi R}}\phi_{+-}^{(2n+1)}\cos \frac{(2n+1) y}{R}~,\\
\phi_{-+}(x^\mu,y) &=&\sum_{n=0}
\frac{1}{\sqrt{\pi R}}\phi_{-+}^{(2n+1)}\sin \frac{(2n+1) y}{R}~,\\
\phi_{--}(x^\mu,y) &=&\sum_{n=0}
\frac{1}{\sqrt{\pi R}}\phi_{--}^{(2n+2)}\sin \frac{(2n+2) y}{R}~.
\ee
Further, three
generations of quarks and leptons are accommodated by
three chiral superfields,
$T^{i}_{10}
=\{U^c~,~Q~,~E^c   \}$ in ${\bf 10}$ and $F^{i}_{\bar{5}}=
\{D^c~,~L    \}$ in $\overline{\bf 5}$,
where $i$ runs over the three generations.
As in \cite{kawamura1}, we assume that they are boundary fields.
To preserve $SU(5)$ symmetry, we have to locate them
at $y=(0,\pi R)$ \cite{hall1}.
Accordingly, the $Z_2\times Z_2'$ invariant Yukawa interactions
can be introduced, and we obtain, after integrating out
the fifth coordinate $y$, the following superpotential \cite{hall1}
\be
W &=&
\frac{Y_{U}^{ij}}{4}\,
T_{10}^{i} T_{10}^{j}H (y=0)+
\sqrt{2}\,Y_D^{ij}\,F_{\bar{5}}^{i}
T_{10}^{j}\overline{H} (y=0)~\nn\\
&=&
\sum_{n=0}~\left[~
\sqrt{2}\,Y_{U}^{ij}\left(~\frac{1}{\sqrt{2^{\delta_{n,0}}}}
Q U^c H_u^{(2n)}+Q Q H_C^{(2n+1)}+U^c E^c
H_C^{(2n+1)}\right)\right.\nn\\
& &\left. +\sqrt{2}\,Y_D^{ij}\left(~
\frac{1}{\sqrt{2^{\delta_{n,0}}}}
Q D^c H_d^{(2n)}+\frac{1}{\sqrt{2^{\delta_{n,0}}}}
L E^c H_d^{(2n)}+Q L H_{\bar{C}}^{(2n+1)}+
U^c D^c H_{\bar{C}}^{(2n+1)}~
\right)~\right]~,
\label{s-potential}
\ee
where $Y_{U}^{ij}$ and $Y_{D}^{ij}$
are the Yukawa couplings.
As we see from (\ref{phi++}), the
normalization of the zero mode $\phi_{++}^{(0)}$
and  the massive $\phi_{++}^{(2n)}$ with $n\neq0$ is different.
Consequently, the normalization
of the Yukawa coupling (\ref{s-potential})
for the zero modes and higher order modes are different.
The factor $\sqrt{2}$ takes care that
only half (not a quarter)  of the original massive Kaluza-Klein modes
are circulating in loops.

The superpotential given by (\ref{s-potential}) admits a $U(1)_R$ symmetry,
which forbids the dimension 5 operators inducing proton decay
\cite{hall1}.
The so-called $\mu$ term is not allowed by the $U(1)_R$ either.
However the D term of $\bar{H} H$ is allowed, offering a possibility that
a $\mu$ term can be produced by the Giudice-Masiero mechanism \cite{giudice}
after local supersymmetry is broken. The $\mu$ term appears via an
explicit or spontaneous breaking of the $U(1)_R$.
So we also consider this case and add
\be
W_{\mu} = \mu_H \bar{H} H
\ee
to the superpotential (\ref{s-potential}).

It is a key point of our setup that the supersymmetry
breaking occurs not only at the branes, but also in the bulk,
which is in contrast to the gaugino mediation \cite{gaugino}
\footnote{Also in ref.~\cite{hall1}, this sort of
supersymmetry breaking is discussed.}.
In the gaugino mediation, the tree level contributions for
the SSB parameters at a fundamental scale $M_{\rm PL}$ are assumed
to be sufficiently suppressed by the sequestering of branes.
However it has been argued \cite{dine} also that such a sequestering
mechanism is not realized in generic supergravity or superstring
inspired models.
In our approach, however, we do not assume the brane
sequestering 
\footnote{If the source of supersymmetry breaking is assumed
only at the hidden sector brane, then the dynamical alignment does not
occur. The point is to assume supersymmetry breaking terms in the
bulk interactions.
}
nor that supersymmetry breaking is flavor
universal at the fundamental scale $M_{\rm PL}$.
So, it may be completely disordered at  $M_{\rm PL}$.

 Further we make the following assumptions on
supersymmetry breaking:
(i) Supersymmetry breaking  does not break
$SU(5)$ gauge symmetry,
(ii) respects five-dimensional Lorentz invariance at least
in the bulk locally,
(iii) is $Z_2 \times Z_2'$ invariant,
(iv) respects  $R$ parity (a part of $U(1)_R$),
(v)  exists also at the brane, and
(vi) appears as soft-supersymmetry breaking terms in the
four-dimensional Lagrangian.
Then the most general Lagrangian of renormalizable
form, in the four dimensional sense,
that satisfies these assumptions is
\be
-L_{\rm SSB} &=&
\left(~\int d^2\theta \eta\int  dy \left\{~
\frac{1}{2}  M W W+ B_H\overline{H} H +B_{H^c}\overline{H}^c H^c +
\frac{1}{2}B_{\Sigma}\Sigma \Sigma
+h_{f}{H} \Sigma H^c
+h_{\bar{f}} \overline{H}^c \Sigma \overline{H}~\right.\right.\nn\\
& &\left.\left.+\frac{\sqrt{2 \pi R}}{2}(\delta(y)-\delta(y-\pi R))(~
\frac{h_{U}^{ij}}{4}\,
T_{10}^{i} T_{10}^{j}H+
\sqrt{2}\,h_D^{ij}\,F_{\bar{5}}^{i}
T_{10}^{j}\overline{H} ~) \right\}+h.c.~\right)\nn\\
& &+\int d^4\theta  \eta \bar{\eta} \int dy \left\{~
\frac{1}{2}m_{\Sigma}^2\Sigma^* \Sigma+
m_{H_d}^2\overline{H}^* \overline{H}+
m_{H_u}^2 H^* H +
m_{H_d^c}^2 \overline{H}^{c*} \overline{H}^c +
m_{H_u^c}^2 H^{c*} H^c \right.\nn\\
& &\left.+\frac{1}{2}(\delta(y)+\delta(y-\pi R))
[~(m_{10}^2)^{i}_{j}T_{10 i}^*T_{10}^{j}+
(m_{5}^2)^{i}_{j}F_{5 i}^*F_{5}^{j}~]~\right\},
\ee
where $\eta = \theta^2$,
$\bar{\eta} = \bar{\theta}^2$ are the external
spurion superfields (which are sometimes interpreted as the
$F$ components of scalar chiral multiplets).

\subsection{Infrared Attractiveness of the SSB Parameters}

We identify
$M_c=1/R$ with $M_{\rm GUT}$ ($\sim 2\times 10^{16}$ GeV), and require
that the MSSM is the effective theory below $M_{\rm GUT}$.
Indeed it has been seen \cite{kubo6, choi1} that sufficiently
strong alignment of the SSB parameters avoiding all the SUSY
flavor problem does not occur in one extra dimensional models,
since the range of energy scale for the power-law running
$M_{\rm PL}/M_c \sim 10^2$ is not large enough.
However it has been also found \cite{protondecay} that the
the compactification scale of the 5D orbifold GUT can be
lowered to $M_c \sim 10^{15}$ GeV with avoiding the rapid
proton decay. 
\footnote{Recently the SSB parameters in a 5D SU(5) GUT model
with one extra dimension compactified on $S^1/Z_2$ \cite{choi2}
have been considered. There also it is reported that
the compactification scale can be lowered and, therefore,
enough alignment in the SSB parameters can be realized.}
Therefore we also assume that
$M_{\rm PL}=10^3 \times M_c$ in this paper.
Then, exactly speaking $M_c$ does not coincide with the gauge
coupling unification scale $M_{\rm GUT}$.
However we neglect this difference of scales as the first approximation
and simply set $M_c=1/R = M_{\rm GUT}$ in calculating the RG flows.

The one-loop $\beta$ functions above $M_{\rm GUT}$ is found to be
($d {\rm A}/d \ln \Lambda = \beta (A)/16\pi^2$,
$N(\Lambda) = X_{\delta=1}R\Lambda/4$):
\be
\beta(g) &=& -8 N~ g^3 ,\\
\beta (Y_t) &=&
\left(-\frac{72}{5}\,g^2+6\,|Y_{t}|^{2}+
4\,|Y_{b}|^{2}\right)\,N~Y_{t},\\
\beta (Y_b) &=& 
\left(-\frac{60}{5}\,g^2+3\,|Y_{t}|^{2}+6\,|Y_{b}|^{2}\right)
\,N~Y_{b},
\ee
where $Y_t = Y_U^{33}$, $Y_b = Y_D^{33}$ and we have neglected
the other elements of the Yukawa couplings. The initial
values of the above couplings are adjusted so as to give their 
low energy values. 
It is noted also that 
the Yukawa couplings $Y_t,~Y_b$ at $M_{\rm GUT}$
 cannot be chosen arbitrarily, because 
$Y_t$ and $Y_b$ are related to the top quark mass $M_t$ and
$\tan\beta=\langle \hat{H} \rangle / \langle \hat{\overline{H}}
\rangle$. So we  use $M_t=174$ GeV
and $M_{\tau}$  (mass of the tau lepton)  $=1.78$ GeV, and impose
the $b-\tau$ unification at
$M_{\rm GUT}$ 
\footnote{ But we will not take the mass of the bottom quark 
very seriously. It becomes  larger than its experimental value.}.

The beta functions for the SSB parameters are given similarly.
Here we introduce the A-parameters for the later purpose by
$h_{U(D)}^{IJ}=-a_{U(D)}^{IJ}Y_{U(D)}^{IJ}~(I,J=1\sim 3)$
and $h_{f(\bar{f})} = -a_{f(\bar{f})} g$.
Then the beta functions for the gaugino mass and the A-parameters
are found to be
\be
\beta (M) &=& -16 N~M~ g^2,\\
\beta (a_t) &=& 
\left( -\frac{192}{5} M g^2+
12 a_t |Y_t|^2+8 a_b |Y_b|^2 +\frac{48}{5}a_f g^2\right) \,
N,\\
\beta (a_b) &=& 
\left( -\frac{168}{5} M g^2+
6 a_t |Y_t|^2+12 a_b |Y_b|^2 +\frac{48}{5}a_{\bar{f}} g^2\right)
\, N,\\
\beta (a_{U}^{ij}) &=&
\left( -\frac{192}{5} M+
\frac{48}{5}a_f \right) \, g^2~N~~~(i,j=1,2),\\
\beta (a_{D}^{ij}) &=&
\left( -\frac{168}{5} M+
\frac{48}{5}a_{\bar{f}} \right)
\, g^2~N~~~(i,j=1,2),\\
\beta (a_f) &=& 
\left(-\frac{196}{5}M +\frac{106}{5} a_f +2 a_{\bar{f}}\right)
\, g^2~N,\\
\beta (a_{\bar{f}}) &=&
\left( -\frac{196}{5}M  +\frac{106}{5} a_{\bar{f}} +2 a_f \right)
\, g^2~N,
\ee
where $a_t = a_U^{33}$, $a_b = a_D^{33}$.

Here we should make some comments on the $\beta$ functions for
$a_D^{i3(3i)}$ and $a_U^{i3(3i)} (i,j=1,2)$.
In the approximation that the Yukawa couplings except for 
$Y_t, Y_b$ are infinitesimally small,
the beta functions for these A-parameters are undetermined.
For example, the beta function for $a_U^{i3}$ is given in
the form of
$Y_U^{i3}\beta(a_U^{i3})=Y_U^{i3}[\cdots]+Y_D^{i3}[\cdots]$.
Therefore, $\beta(a_U^{i3})$ becomes dependent on the ratio of 
$Y_D^{i3}/Y_U^{i3}$. We may evaluate these small Yukawa couplings
and give the beta functions. Here, however, we take another way.
Note that we may obtain $\beta (a_D^{i3})$ unambiguously as
\be
\beta (a_D^{i3}) &=&
\left( 4 a_D^{i3} Y_b^2-\frac{168}{5} M g^2  +
6 a_t |Y_t|^2+8 a_b |Y_b|^2 +
\frac{48}{5}a_{\bar{f }}g^2\right) \,N~~~(i=1,2).
\ee
On the other hand, it will be seen 
that only the approximate values of the  A-parameters at 
$M_{\rm GUT}$ are important for the later arguments.
Hence we estimate $a_D^{3i}$ and $a_U^{i3(3i)}$ at $M_{\rm GUT}$
simply by substituting them with $a_D^{i3}$.

For the soft scalar masses, the beta functions are given as
follows; 
\be
\beta (m_{H_d}^{2}) &=&
\left( -\frac{96}{5} g^2 |M|^{2}+\frac{48}{5}g^{2}
 (|a_{\bar{f}}|^2+m_{H_d^c}^{2}+
m_{H_d}^{2}+m_{\Sigma}^{2})\right) N,
\label{betamd2}\\
\beta (m_{H_d^c}^{2}) &=&
\left( -\frac{96}{5} g^2 |M|^{2}+\frac{48}{5}g^{2}
 (|a_{\bar{f}}|^2+m_{H_d^c}^{2}+
m_{H_d}^{2}+m_{\Sigma}^{2})\right) N,\\
\beta (m_{H_u}^{2}) &=&
 \left(-\frac{96}{5}g^2 |M|^{2}+\frac{48}{5}g^{2}(|a_f|^2
 +m_{H_u}^{2}+
m_{H_u^c}^{2}+m_{\Sigma}^{2})\right)N,
\label{betamu2}\\
\beta (m_{H_u^c}^{2}) &=&
 \left(-\frac{96}{5}g^2 |M|^{2}+\frac{48}{5}g^{2}(|a_f|^2
 +m_{H_u}^{2}+
m_{H_u^c}^{2}+m_{\Sigma}^{2})\right)N,\\
\beta (m_{\Sigma}^{2}) &=&
 \left( -40 g^2|M|^{2}+2g^{2} (|a_{\bar{f}}|^2+m_{H_d^c}^{2}+
m_{H_d}^{2}+m_{\Sigma}^{2})\right.\nn\\
& +&\Bigl. 2g^{2} (|a_{f}|^2+m_{H_u^c}^{2}+
m_{H_u}^{2}+m_{\Sigma}^{2})\Bigr) N,\\
\beta (m_{5^{33}}^{2}) &=&
 \left(-\frac{96}{5} g^2 |M|^{2}+
8 |Y_{b}|^{2}(
m_{H_d}^{2}+m_{10^{33}}^{2}+m_{5^{33}}^{2})+
8 |a_{b}Y_b|^{2} \right) N,
\label{3stm1}\\
\beta (m_{10^{33}}^{2}) &=&
 \left(-\frac{144}{5} g^2 |M|^{2}+6|Y_{t}|^{2}(
m_{H_u}^{2}+2m_{10^{33}}^{2})+
4 |Y_{b}|^{2}(
m_{H_d}^{2}+m_{10^{33}}^{2}+m_{5^{33}}^{2})\right.\nn\\
 &+ & \biggl. 6|a_{t} Y_t|^{2}+4  |a_{b}Y_b|^{2}\biggr) N,
\label{3stm2}\\
\beta (m_{5^{i3}}^{2}) &=& 4 m_{5^{i3}}^{2} |Y_b|^2 N
~~~(i=1,2),\\
\beta (m_{10^{i3}}^{2}) &=&
m_{10^{i3}}^{2}(3 |Y_t|^2+2 |Y_b|^2)N~~~(i=1,2),
\label{13stm}\\
\beta (m_{5^{ij}}^{2}) &=&
-\frac{96}{5} g^2 |M|^{2}\delta_{ij}N~~~(i,j=1,2),\\
\beta (m_{10^{ij}}^{2}) &= &
-\frac{144}{5} g^2 |M|^{2}\delta_{ij}N~~~(i,j=1,2).
\label{1stm}
\ee

If we neglect the Yukawa couplings in the beta functions, then
it is found that both of $a/M$ and $m^2/M^2$ rapidly approaches to
their ``infra-red fixed point" values, 
which are flavor universal \cite{kubo6}.
The rate of convergence of the SSB parameters from the Planck
scale to the GUT scale is roughly given by
$g^2(M_{\rm GUT})/g^2(M_{\rm PL})$ for the A-parameters and
$(g^2(M_{\rm GUT})/g^2(M_{\rm PL}))^2$ for the squared soft scalar masses.
Contrary to the K-K GUT models, however, radiative corrections
by the Yukawa interactions also show power-law behavior.
Therefore, the (3,3) component of the A-parameters and the
soft scalar masses converge to values different from the
(1,1) or (2,2) components. This discrepancy may cause
new sources of FCNC compared with the case that only the
gauge multiplets propagate in the bulk \cite{kubo6}.

Now we evaluate the actual converging behavior of the SSB
parameters at $M_{\rm GUT}$.
To be explicit, in what follows we consider only one case:
$g = (0.0406\times 4\pi)^{1/2},
M_{\rm GUT} =1.83 \times 10^{16}~\mbox{GeV}$, and
\be
Y_t= 0.899 g,~
Y_b= 0.0397 g,
~\tan\beta=3.7~.
\label{case2}
\ee
Note that the mass scale of the SSB parameters are totally
determined by the GUT gaugino mass due to convergence towards
the infra-red fixed points. So we evaluate the SSB parameters
in the unit of the gaugino mass at $M_{\rm GUT}$, which may
be chosen freely.

The RG flows between $M_{\rm PL}$ and $M_{\rm GUT}$
for $m^2_{10}$, $m^2_{5}$, Re$[a_U]$, Re$[a_D]$ are shown
in Figs.1, 2, 3 and 4, respectively.
\begin{figure}[htb]
\includegraphics*[width=0.6\textwidth]{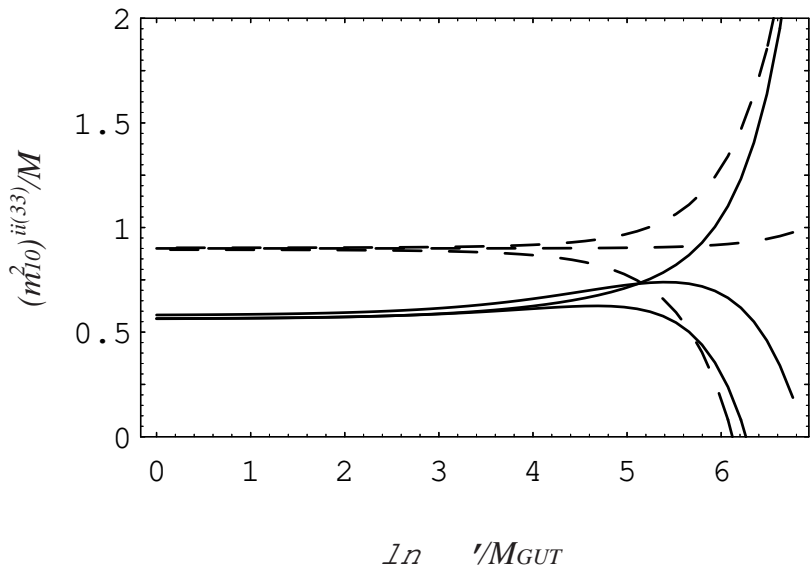}
\caption{\label{fig1} Converging RG flows of 
${m_{10}^2}^{ii}/M^2 (i=1,2)$(dashed line) and 
${m_{10}^2}^{33}/M^2$(solid line).}
\end{figure}
\begin{figure}
\includegraphics*[width=0.6\textwidth]{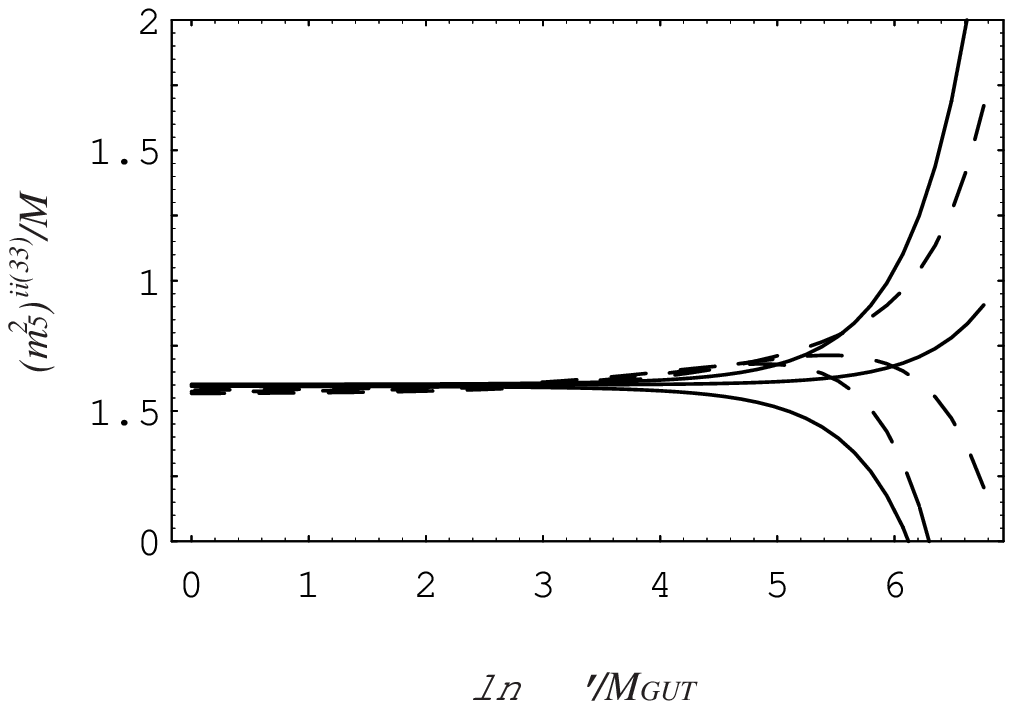}
\caption{\label{fig2} Converging RG flows of 
${m_{5}^2}^{ii}/M^2 (i=1,2)$(dashed line) and 
${m_{5}^2}^{33}/M^2$(solid line).}
\end{figure}
\begin{figure}
\includegraphics*[width=0.6\textwidth]{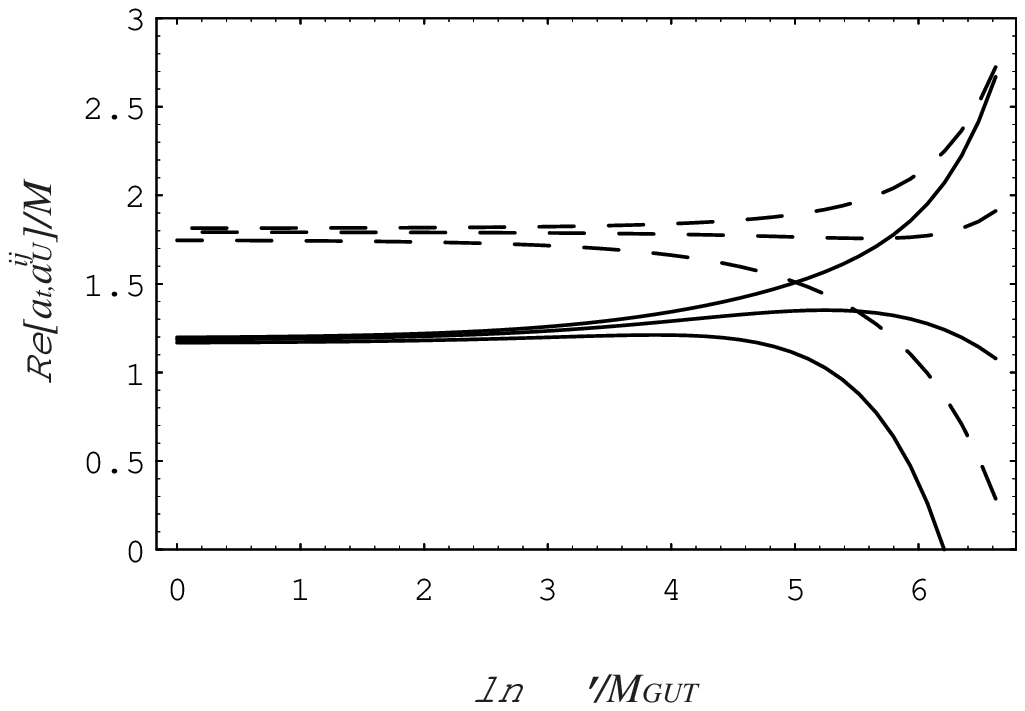}
\caption{\label{fig3} Converging RG flows of 
$\mbox{Re}[a^{ij}_{U}]/M (i,j=1,2)$(dashed line) and 
$\mbox{Re}[a_t]/M$(solid line).}
\end{figure}
\begin{figure}
\includegraphics*[width=0.6\textwidth]{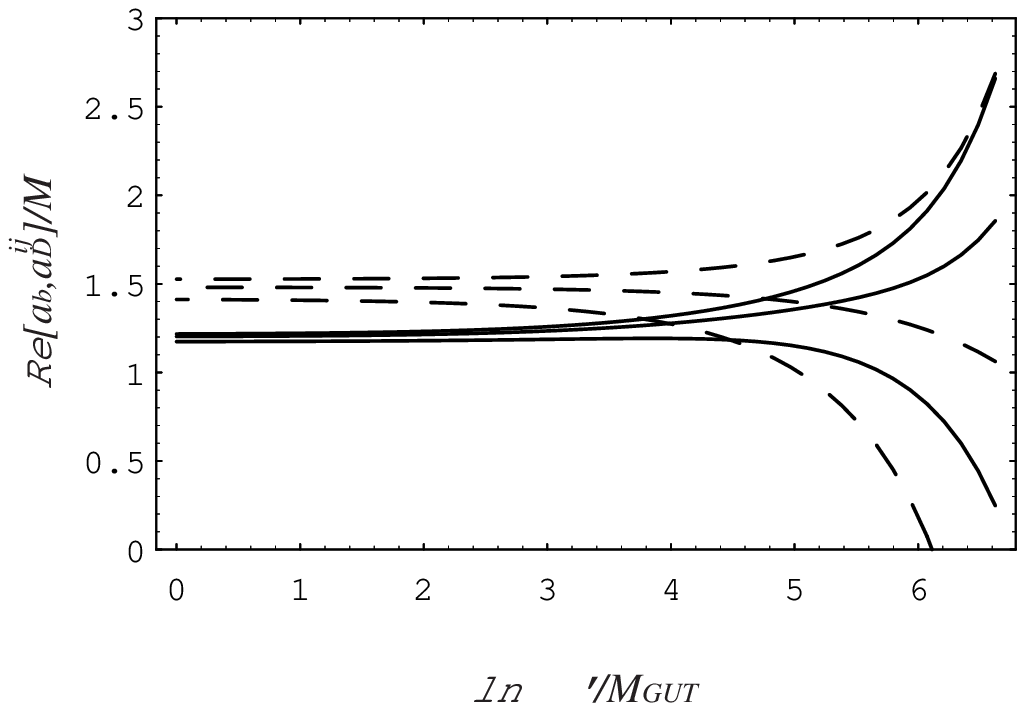}
\caption{\label{fig4} Converging RG flows of 
$\mbox{Re}[a^{ij}_{D}]/M (i,j=1,2)$(dashed line) and 
$\mbox{Re}[a_b]/M$(solid line).}
\end{figure}
%\begin{figure}
%\includegraphics{m10ii.eps}
%\caption{\label{fig:wide} Infrared attractiveness of
%${m_{10}^2}^{ii}/M^2$.}
%\end{figure}
%\begin{figure}
%\includegraphics{m5ii.eps}
%\caption{\label{fig:wide} Infrared attractiveness of
%${m_{5}^2}^{ii}/M^2$.}
%%\end{figure}
%\begin{figure}
%%\includegraphics{m10i3.eps}
%\caption{\label{fig:wide} Infrared attractiveness of
%$\mbox{Re}{m_{10}^2}^{i3}/M^2$.}
%\end{figure}
%\begin{figure}
%%\includegraphics{mhu.eps}
%\caption{\label{fig:wide} Infrared attractiveness of
%${m_{H_u}^2}/M^2$.}
%\end{figure}
%\begin{figure}
%\includegraphics{msigma.eps}
%\caption{\label{fig:wide} Infrared attractiveness of
%${m_{\Sigma}^2}/M^2$.}
%\end{figure}
%\begin{figure}
%\includegraphics{reauij.eps}
%\caption{\label{fig:wide} Infrared attractiveness of
%$\mbox{Re}a_U^{ij}/M$.}
%\end{figure}
%\begin{figure}
%%%\includegraphics{readi3.eps}
%\caption{\label{fig:wide} Infrared attractiveness of
%$\mbox{Re}a_D^{i3}/M$.}
%\end{figure}
%\begin{figure}
%\includegraphics{imat.eps}
%\caption{\label{fig:wide} Infrared attractiveness of
%$\mbox{Im}a_t/M$.}
%\end{figure}
%\begin{figure}
%%\includegraphics{imab.eps}
%\caption{\label{fig:wide} Infrared attractiveness of
%$\mbox{Im}a_b/M$.}
%\end{figure}
%\begin{figure}
%\includegraphics{imadij.eps}
%\caption{\label{fig:wide} Infrared attractiveness of
%$\mbox{Im}a_D^{ij}/M$.}
%\end{figure}
%\begin{figure}
%%\includegraphics{af.eps}
%\caption{\label{fig:wide} Infrared attractiveness of
%$\mbox{Re}a_f/M$.}
%\end{figure}
%%\begin{figure}
%\includegraphics{b.eps}
%\caption{\label{fig:wide} Infrared attractiveness of
%$\mbox{Re}b_H/M$.}
%\end{figure}
It is seen that the convergence of the squared soft scalar masses
are remarkable. Also the discrepances of the converging values
between $(m^2)^{ii} (i=1,2)$ and $(m^2)^{33}$ are sizable for
the 10-multiplets. The convergence of the A-parameters are weak
compared with the soft scalar masses. However it will be seen that
this degree of convergence gives quite enough alignment satisfying
the FCNC bounds. Rather what we have to care is the sizable
discrepancies between $a^{ij} (i,j=1,2)$ and $a^{33}$.

To be more explicit, we also give the converging values of SSB 
parameters at $M_{\rm GUT}$ in Tables 1 and 2. The range of convergence
is evaluated by starting with  the initial values of
$m^2/M^2 \in [-1,1]$ for the soft scalar masses and 
$a/M \in [-1,1]$ for the A-parameters at $M_{\rm PL}$.
The infra-red fixed point values in the case of fixed 
(non-running) Yukawa couplings are also shown.

\begin{table}[htb]
\caption{\label{table1}The converging values of the squared soft 
scalar masses at $M_{\rm GUT}$ starting with the initial values of
$m^2/M^2 \in [-1,1]$ at $M_{\rm PL}$. The indeces run $i,j=1,2$.
Their infra-red fixed point values in the case of fixed 
(non-running) Yukawa couplings are also shown.}
\begin{center}
\begin{tabular}{|c|c|c|}\hline
 & ~~IR fixed points ~~& ~~Convergence at $M_{\rm GUT}$~~ \\ \hline
$\left( m_{10}^2 \right)^{33}/M^2$ & 0.54 & 0.574 $\pm$0.005 \\
$\left( m_{5}^2 \right)^{33}/M^2$ & 0.60 & 0.598 $\pm$0.002 \\
$\left( m_{10}^2 \right)^{ii}/M^2$ & 0.90 & 0.899 $\pm$0.001 \\
$\left( m_{5}^2 \right)^{ii}/M^2$ & 0.60 & 0.599 $\pm$ 0.001 \\
$\mbox{Re}\left[\left(m_{10(5)}^2 \right)^{ij}\right]/M^2$ & 0 &
0 $\pm$0.002 \\
$\mbox{Im}\left[\left(m_{10(5)}^2 \right)^{ij}\right]/M^2$ & 0 &
0 $\pm$0.002 \\
$\mbox{Re}\left[\left( m_{10(5)}^2 \right)^{i3}\right]/M^2$ & 0 &
0 $\pm 0.001( 0.002)$\\
$\mbox{Im}\left[\left( m_{10(5)}^2 \right)^{i3}\right]/M^2$ & 0 &
0 $\pm 0.001( 0.002)$\\
$m_{h_u}^2/M^2$ & 0 & 0.0016 $\pm$ 0.0004 \\
$m_{h_d}^2/M^2$ & 0 & 0.0012 $\pm$ 0.0008 \\
$m_\Sigma^2/M^2$ & 1 & 0.9991 $\pm$ 0.0008 \\ \hline
\end{tabular}
\end{center}
\end{table}
%\nopagebreak[4]
\begin{table}
\caption{\label{table2}The converging values of the A-parameters 
at $M_{\rm GUT}$ starting with the initial values of
$a/M \in [-1,1]$ at $M_{\rm PL}$. The indeces run $i,j=1,2$.
Their infra-red fixed point values in the case of fixed 
(non-running) Yukawa couplings are also shown.}
\begin{center}
\begin{tabular}{|c|c|c|}\hline
 & ~~IR fixed points~~ & ~~Convergence at $M_{\rm GUT}$~~ \\ \hline
$\mbox{Re}\left[a_t\right]/M$ & 1.12 & 1.18$\pm$0.01 \\
$\mbox{Re}\left[a_b\right]/M$ & 1.26 & 1.17$\pm$0.01 \\
$\mbox{Re}\left[a_U^{ij}\right]/M$ & 1.80 & 1.75$\pm$0.02 \\
$\mbox{Re}\left[a_D^{ij}\right]/M$ & 1.50 & 1.46$\pm$0.02 \\
$\mbox{Re}\left[a_D^{i3}\right]/M$ & 1.16 & 1.17$\pm$0.01 \\
$\mbox{Im}\left[a_t\right]/M$ & 0 & 0$\pm$0.01 \\
$\mbox{Im}\left[a_b\right]/M$ & 0 & 0$\pm$0.02 \\
$\mbox{Im}\left[a_{U,D}^{ij}\right]/M$ & 0 & 0$\pm$0.02 \\
$\mbox{Im}\left[a_D^{i3}\right]/M$ & 0 & 0$\pm$0.02 \\
$\mbox{Re}\left[a_f \right]/M$ & 1 & 0.98988$\pm$0.00007  \\
%${\hat b}/M$ & 0 & -0.047$\pm$0.065\\ 
\hline
\end{tabular}
\end{center}
\end{table}

Lastly let us make some remarks on the parameters $\mu_H$ and $B_H$.
It should be noted that $\mu_H$ receives only logarithmic
corrections due to the $N=2$ supersymmetry in the bulk.
On the other hand the soft parameter $B_H$ does not show converging
behavior, since its beta function, which is given explicitly as
\be
\beta (B_H) =
\left( \frac{96}{5}Mg^2-\frac{48}{5}a_f g^2
-\frac{48}{5}a_{\bar{f}} g^2\right) N,
\ee
vanishes rapidly with scaling down.
Thus we cannot explain the $\mu$ term by the RG running behavior
in the extra-dimensions. However it has been known also that the
$\mu$ and $B$ parameters may be generated at lower energy scale
by assuming extra scalar fields \cite{hall1, muterm}.
Therefore we suppose the $\mu$ term to be generated by other
mechanisms at lower energy scale and take $\mu_H$ and
$B_H$ as free parameters in our analysis.

\section{Evaluation of FCNCs and CP violations at $M_{\rm SUSY}$}
%$\delta_{LL}$, $\delta_{RR}$ and $\delta_{LR}$}
We are now interested in verifying whether the
infrared attractive values of the SSB parameters 
given in TABLE I and II are consistent with the experimental
constraints coming from the dangerous FCNC
and CP-violating processes at low energies.
For this purpose, first the SSB parameters should be evaluated
at $M_{\rm SUSY}$.
We operate the two-loop RG functions to calculate
the low-energy values of the dimensionless parameters,
while we use  the one-loop RG functions for the SSB
parameters.
Then the flavor mixing masses, which are of our present
concern in evaluating the amount of FCNC, are evaluated
in the bases of the mass eigenstates for the quarks and 
leptons.
To begin with, we recall that
the mass matrices ${\bf M}_{I}~(I=u,d,e)$ for the quarks and leptons
and ${\bf M}_{\nu}$ for the left-handed neutrinos, respectively,
are diagonalized by the unitary matrices as
\be
U_{u(d,e)L}^{\dag}{\bf M}_{u(d,e)}U_{u(d,e)R}
&=&\mbox{diag} (m_{u(d,e)}, m_{c(s,\mu)},m_{t(b,\tau)}),
\label{UeL}\\
U_{\nu}^{T}{\bf M_\nu}U_{\nu} &=&
\mbox{diag} (m_{\nu_e},m_{\nu_\mu},m_{\nu_\tau}).
\label{Unu}
\ee
These diagonalization matrices are not known, unless the
matrices of Yukawa couplings are explicitly fixed.
However, the mixing matrices defined by
\be
V_{CKM} &=&U_{uL}^{\dag} U_{dL}~,~
V_{MNS} =U_{eL}^{\dag} U_{\nu},
\label{ckm1}
\ee
are observables. Roughly $V_{\rm CKM}$ and $V_{\rm MNS}$
may be represented by the following matrices,
\be
V_{\rm CKM} &\sim &\left(
\begin{array}{ccc}
0.98  &  0.22 & 0.003 \\
-0.22  & 0.97  & 0.04 \\
0.01  & -0.04  & 1 \\
\end{array}\right)~,~~~~
V_{\rm MNS} =\left(
\begin{array}{ccc}
1/ \sqrt{2}  & 1/ \sqrt{2}  & 0 \\
-1/2  & 1/2  & 1/ \sqrt{2} \\
1/2  & -1/2  & 1/ \sqrt{2} \\
\end{array}\right).
\label{mixings}
\ee 
So we perform order estimation of the flavor mixing masses
by simply assuming that mixings of the diagonalization 
matrices are similar to those of the above mixing matrices.

\subsection{The slepton sector}
First we consider the soft scalar masses of sleptons.
The slepton mass matrices at $M_{\rm SUSY}$  are found to be
\be
\frac{{\bf \tilde{m}^2}_{LL}}{M^2}& =  &
%m^2_{\tilde{\ell}}
\left( 
\begin{array}{ccc}
1.152 \pm 0.002&0\pm 0.002(1+I) & 0\pm 0.002(1+I)  \\
0\pm 0.002(1+I)  & 1.152 \pm 0.002
& 0\pm 0.002(1+I)  \\
 0\pm 0.002(1+I)  & 0\pm 0.002(1+I)
 &1.148 \pm 0.002 
\end{array}
\right),\nn \\
\frac{{\bf \tilde{m}^2}_{RR}}{M^2} &=&
%m^2_{\tilde{\ell}}
\left( 
\begin{array}{ccc}
1.054 \pm 0.002 &0\pm 0.002(1+I)
& 0\pm 0.001(1+I)  \\
0\pm 0.002(1+I)  & 1.054 \pm 0.002
& 0\pm 0.001(1+I)\\
0\pm 0.001(1+I) & 0\pm 0.001(1+I)
&0.718\pm 0.005  
\end{array}
\right),
\label{scalarmass} 
\ee
where $M$ denotes the gaugino mass at $M_{\rm GUT}$.
Note that the branching ratios of  lepton flavor violating processes
are proportional to the off-diagonal elements of
$U_{eL}^{\dagger} {\bf \tilde{m}^2}_{LL} U_{eL}$
and $U_{eR}^{\dagger} {\bf \tilde{m}^2}_{RR} U_{eR}$ \cite{fcnc},
where
 the unitary matrices
$U_{eL}$ and $U_{eR}$ are not explicitly known.
In our following calculations, we first assume that the 
rotation matrices to be
\be
U_{eR} \sim V_{\rm CKM}~,~~~U_{eL} \sim V_{\rm MNS}^{\dag},
\label{eReL}
\ee
which are regarded as their maximal estimations.
According to \cite{fcnc}, we then calculate the ratios of the
off-diagonal elements of
$U_{eL(R)}^{\dagger}{\bf \tilde{m}^2}_{LL(RR)} U_{eL(R)}$
to their diagonal elements $(m_{\tilde l}^2)$, which are 
denoted by $(\delta^{\ell}_{ij})_{LL(RR)}$.

Here the origin of FCNC can be separated into two parts.
(i)The difference of the fixed point for $({\bf \tilde{m}^2})^{ii}(i=1,2)$
and $({\bf \tilde{m}^2})^{33}$ because of $Y_{t,b}$.
Note that ${\bf \tilde{m}^2}_{LL}$ and ${\bf \tilde{m}^2}_{RR}$ are 
embedded in the $m_5^2$ and $m_{10}^2$ respectively, 
and that the $\beta$ function of
$m_5^2$ depends only on $Y_b$ while that of $m_{10}^2$ on
$Y_t$ ((\ref{3stm1})$\sim$(\ref{1stm})). 
Since these effects of $Y_{t,b}$
spoil the degeneracy for the soft scalar masses, the
flavor mixing masses generating FCNC arise through the 
rotation given  by Eq.~(\ref{eReL}).
(ii)The deviation from the fixed points also can be the
origin of FCNC. Since each parameter cannot converge exactly 
to the IR fixed point due to finite energy range of the GUT
theory, there are small deviations from the fixed points.
Therefore if we assume the initial value of parameters at 
$M_{\rm PL}$ to be arbitrary, then misalignment of the soft masses
remains slightly and also generates FCNC.

However, it is found that ${\bf \tilde{m}^2}_{LL(RR)}$ given
above are enough degenerate so as to suppress FCNC. 
While ${\bf \tilde{m}^2}_{RR}$ is strongly affected
by $Y_t$ (in the meaning of (i)), the discrepancy does not
give rise to a large contribution to the off-diagonal elements
of $\delta_{RR}$ because of the small mixing matrix $V_{\rm CKM}$. 
On the other hand, the degeneracy in ${\bf \tilde{m}^2}_{LL}$ is 
good enough, even if it is transformed by
the large mixing matrix $V_{\rm MNS}$ (\ref{eReL}).
This is one of our main findings on the orbifold GUT model.

Then, estimating $\delta$'s by taking into account the effects both
(i) and (ii), we find:
\be
\mbox{Re}\left[ (\delta^{\ell}_{12})_{LL}\right]
&\sim &  3.1 \times 10^{-3}~ ,~
\mbox{Im}\left[ (\delta^{\ell}_{12})_{LL}\right]
\sim 2.0 \times 10^{-3}
%\left( \frac{100}{m_{\tilde{\ell}}(\mbox{GeV})}\right)^2
\label{lep12LL},\\
\mbox{Re}\left[(\delta^{\ell}_{13})_{LL}\right]
&\sim & 3.8 \times 10^{-3}~,~
\mbox{Im}\left[(\delta^{\ell}_{13})_{LL}\right]
\sim 2.5 \times 10^{-3}~
%\left( \frac{100}{m_{\tilde{\ell}}(\mbox{GeV})}\right)^2
 \label{lep13LL},\\
\mbox{Re}\left[(\delta^{\ell}_{23})_{LL}\right]
&\sim & 3.8 \times 10^{-3} ~,~
\mbox{Im}\left[(\delta^{\ell}_{23})_{LL}\right]
\sim 2.5 \times 10^{-3}
%\left( \frac{100}{m_{\tilde{\ell}}(\mbox{GeV})}\right)^2
\label{lep23LL},\\ 
%(\delta^{\ell}_{12})_{RR}
%&\sim& 8.28 \times 10^{-8}, \\
%(\delta^{\ell}_{13})_{RR}
%&\sim& - 5.49 \times 10^{-2}, \\
%(\delta^{\ell}_{23})_{RR}
%&\sim& - 1.39 \times 10^{-6}.
\mbox{Re}\left[ (\delta^{\ell}_{12})_{RR}\right]
&\sim &  2.1 \times 10^{-3}~ ,~
\mbox{Im}\left[ (\delta^{\ell}_{12})_{RR}\right]
\sim 1.5\times 10^{-3}~
%\left( \frac{100}{m_{\tilde{\ell}}(\mbox{GeV})}\right)^2
\label{lep12RR},\\
\mbox{Re}\left[(\delta^{\ell}_{13})_{RR}\right]
&\sim & 3.0 \times 10^{-3}~,~
\mbox{Im}\left[(\delta^{\ell}_{13})_{RR}\right]
\sim 1.3 \times 10^{-3}~
%\left( \frac{100}{m_{\tilde{\ell}}(\mbox{GeV})}\right)^2
 \label{lep13RR},\\
\mbox{Re}\left[(\delta^{\ell}_{23})_{RR}\right]
&\sim & 1.6 \times 10^{-2}~,~
\mbox{Im}\left[(\delta^{\ell}_{23})_{RR}\right]
\sim 1.3 \times 10^{-3}~
%\left( \frac{100}{m_{\tilde{\ell}}(\mbox{GeV})}\right)^2
\label{lep23RR}.
\ee
The experimental upper bounds of $(\delta^{\ell}_{ij})_{LL,RR}$
are given 
in \cite{fcnc}, and  are shown in TABLE~III.
Constraints appearing in TABLE~III are shown in the case that 
the ratio of the squared photino mass and the squared slepton mass
is 0.3. Actually in the case of (\ref{case2}),~
the gaugino and the average sfermion masses are found to be
\be 
m_{{\tilde q}_{u(d)}}=3.03(3.12) M, ~~
m_{\tilde l}=1.02 M,~~
m_{\tilde{\gamma}} =  M_1 = 0.39 M,~~
m_{\tilde{g}} = M_3 = 3.36 M
\label{ratio}
\ee 
at the weak scale. 
Therefore the ratio 
$m^2_{\tilde{\gamma}}/m^2_{\tilde{\ell}}$ is about $0.15$.
The constraints for this case are not much different from those
given in TABLE~III.
Here it should be noted also that the above ratios are
low energy predictions for the orbifold GUT model, which
are completely independent of the fundamental physics at
$M_{\rm PL}$.
Comparing the results given
in (\ref{lep12LL})--(\ref{lep23RR}) with TABLE~III,
it is  seen that the off-diagonal elements
$(\delta^{\ell}_{ij})_{LL,RR}$
are small enough to satisfy the constraints.

\begin{table}[thb]
\begin{center}
\caption{Limits on the $|\delta^{\ell}_{ij}|$ from
$\ell_j \rightarrow \ell_i \gamma$ decays,EDM of the
electron for 
$m^2_{\tilde{\gamma}}/m^2_{\tilde{\ell}} = 0.3$.
Here the parameter $\tilde{m}_{\tilde{\ell}}$ denotes
$m_{\tilde{\ell}}(\mbox{GeV})/100$.
See \cite{fcnc} for details.}
\begin{tabular}{|c|c|c|} \hline \hline
$|(\delta^{\ell}_{12})_{LL,RR}|$ & $|(\delta^{\ell}_{13})_{LL,RR}|$
&$|(\delta^{\ell}_{23})_{LL,RR}|$
\\ \hline
$ 4.1 \times 10^{-3} ~\tilde{m}_{\tilde{\ell}}^2 $
& $ 1.5 \times 10^{+1} ~\tilde{m}_{\tilde{\ell}}^2 $
& $ 2.8  ~\tilde{m}_{\tilde{\ell}}^2$
\\ \hline \hline
$|(\delta^{\ell}_{12})_{LR}|$ & $|(\delta^{\ell}_{13})_{LR}|$
&$|(\delta^{\ell}_{23})_{LR}|$
\\ \hline
$ 1.4 \times 10^{-6} ~ \tilde{m}_{\tilde{\ell}}^2 $
& $ 8.9 \times 10^{-2} ~ \tilde{m}_{\tilde{\ell}}^2 $
& $ 1.7 \times 10^{-2} ~\tilde{m}_{\tilde{\ell}}^2$
\\ \hline \hline
%$|\mbox{Re}(\delta^{\ell}_{11})_{LR}|$
%& $|\mbox{Re}(\delta^{\ell}_{22})_{LR}|$
%& $|\mbox{Re}(\delta^{\ell}_{33})_{LR}|$
%\\ \hline 
%$ 1.1 \times 10^{-2}  ~ \tilde{m}_{\tilde{\ell}}^{-1} $
%& $ 2.1 ~ \tilde{m}_{\tilde{\ell}}^{-1} $
%& $ 3.6 \times 10^{+1}~\tilde{m}_{\tilde{\ell}}^{-1}$
%\\ \hline \hline
%$|\mbox{Im}(\delta^{\ell}_{11})_{LR}|$
%& 
%&
%\\ \hline 
% $ 3.0 \times 10^{-7}~\tilde{m}_{\tilde{\ell}}$
%&
%&
%\\ \hline \hline
\end{tabular}
\end{center}
\end{table}

%The sfermion and gaugino masses in (\ref{ratio}) might be so large
%that FCNC is suppressed by the decoupling mechanism. However, in our
%calculation, soft scalar masses are determined only by the gaugino mass
%at $M_{GUT}$ and the values in (\ref{scalarmass}) are unchanged by
%the choice of the gaugino mass $M$. Hence, $\delta$'s
%in (\ref{lep12LL})$\sim$(\ref{lep23RR}) do not depend on $M$ and
%$m_{\tilde{\ell}}$.
%\footnote{Eqs.(\ref{scalarmass}) show that diagonal elements of
%${\bf \tilde{m}^2}_{LL}$
% are more degenerated than that of ${\bf  \tilde{m}^2}_{RR}$.
% However, $\delta $'s
% are the same order because of deviations
% from IR fixed points of off-diagonal elements. If let the off-diagonal
% elements be exactly zero, order of $\delta$'s are
% $(\delta^{\ell})_{LL}\sim 10^{-(2 \sim 3)}~\tilde{m}_{\tilde{\ell}}^{-2}$and 
% $(\delta^{\ell})_{RR}\sim 10^{(0 \sim-2)}~\tilde{m}_{\tilde{\ell}}^{-2}$.}

To satisfy the FCNC constraints,
it is also necessary to take into account the mass matrices 
among the left-handed and right-handed sleptons, which
are generated through the A-terms .
The A-parameters $a_e$ at the weak scale are found to be
\be
\frac{a_e}{M}=\left(
\begin{array}{ccc}
a & a & b \\
a & a & b \\
b & b & c 
\end{array}\right),~~~
\begin{array}{l}
a=(2.175 \pm 0.023)+I~(0 \pm 0.023) \\
b=c=(1.888 \pm 0.014)+I~(0 \pm 0.018).
\end{array}
\label{ae}
\ee
The left-right mixing mass matrix is given by
${\bf \tilde{m}^2}_{LR}=v_d h_e=-v_d(Y_e a_e)$, where
\be 
v_d Y_e=U_{eL}~m_e^{diag}~U_{eR}^{\dag}=
\left( \begin{array}{ccc}
 -6.9 & -12.1 & 890.0 \\
 7.7 & 11.9 & -890.0 \\
 22.9 & 128.6 & 1252.0
\end{array} \right) ~(\mbox{MeV}).
\label{leptonmass}
\ee

There are two origins of FCNC here again, that is, (i)the flavor
non-universality in the fixed points due to the Yukawa couplings. 
It spoils the alignment of the A-term. 
(ii)The deviations from the fixed point values similar to 
the case of the scalar masses. This effect is actually irrelevant
for the left-right mixing masses.
Now the index $(\delta^{\ell})_{LR}$  is given by the ratio
of ${\bf \tilde{m}^2}_{LR}$ and the average slepton mass 
squared $m^2_{\tilde{\ell}}$. 
Note that the left-right mixing mass matrix is given by
the product of the A-parameters proportional to the
gaugino mass scale and the lepton mass matrix (\ref{leptonmass}),
which does not depend on the gaugino mass. Therefore, $\delta_{LR}$
is dependent on the SSB mass scale differently from
$\delta_{LL(RR)}$.
Here we represent the index in the unit of $100{\mbox{GeV}}$
for the slepton mass.
Then $(\delta^{\ell})_{LR}$ are estimated as follows,
\be
\mbox{Re}[(\delta^{\ell}_{12})_{LR}]
&\sim& 1.3\times 10^{-4}
\left( \frac{100}{m_{\tilde{\ell}}(\mbox{GeV})}\right)^2,~
\label{deltaLR12}\\
\mbox{Re}[(\delta^{\ell}_{13})_{LR}]
&\sim& 1.3\times 10^{-4}
\left( \frac{100}{m_{\tilde{\ell}}(\mbox{GeV})}\right)^2,~
\label{deltaLR13}\\
\mbox{Re}[(\delta^{\ell}_{23})_{LR}]
&\sim& 3.8\times 10^{-4}
\left( \frac{100}{m_{\tilde{\ell}}(\mbox{GeV})}\right)^2.
\label{deltaLR23}
\ee
All the fixed points are real and, therefore, the imaginary parts
 of the left-right mixings can be treated as zero.

Unfortunately $(\delta^{\ell}_{12})_{LR}$ obtained in this 
analysis exceeds the experimental bound given in TABLE III.
This is due to the large mixings of the $V_{\rm MNS}$ matrix.
\footnote{In practice, $(\delta^{\ell}_{12})_{LR}$ turns
out to be smaller than the bound as long as we use the
bi-maximal form of the $V_{\rm MNS}$ matrix given by (\ref{mixings}).
However the more viable form of $V_{\rm MNS}$ matrix gives
a sizable contribution. Anyway $(\delta^{\ell}_{21})_{LR}$ exceeds
the bound.}
%$(U_{eL}^{\dag})_{11}(h_e)_{1I}(U_{eR})_{I2}$
%and $(U_{eL}^{\dag})_{12}(h_e)_{2I}(U_{eR})_{I2}~(I=1 \sim 3)$
%are nearly canceled for each $I$,
%and there are no contributions from
%$(U_{eL}^{\dag})_{13}(=(V_{MNS})_{13}=0)$.
%The value in (\ref{deltaLR12}) is a larger one of 
%$(\delta^{\ell}_{12})_{LR}=
%v_d(U_{eL}^{\dag} h_e U_{eR})_{12}/m^2_{\tilde{\ell}}$ and
%$(\delta^{\ell}_{21})_{LR}
%=v_d(U_{eL}^{\dag} h_e U_{eR})_{21}/m^2_{\tilde{\ell}}$.
%Calculating $(\delta^{\ell}_{12})_{LR}$, contributions from
%$(U_{eL}^{\dag})_{11}(h_e)_{1I}(U_{eR})_{I2}$
%and $(U_{eL}^{\dag})_{12}(h_e)_{2I}(U_{eR})_{I2}~(I=1 \sim 3)$
%are nearly canceled for each $I$,
%and there are no contributions from
%$(U_{eL}^{\dag})_{13}(=(V_{MNS})_{13}=0)$.
%On the other hand for $(\delta^{\ell}_{21})_{LR}$, such a cancellation
%does not occur and there exist contributions from
%$(U_{eL}^{\dag})_{23}$.
%Then, $(\delta^{\ell}_{21})_{LR}$ gets the large value because of
%the form of $V_{MNS}$.
However, this implies only that our assumption $U_{eL}=V_{\rm MNS}$,
which would be maximal estimation of the matrix, is not
viable phenomenologically in the orbifold GUT models.
Therefore let us repeat the above estimation by 
setting $U_{eL}=V_{\rm CKM}$, for example.
Then it is found that the indices $(\delta_{ij})_{LR}$ become 
fairly smaller, such as
\be
\mbox{Re}[(\delta^{\ell}_{12})_{LR}]
&\sim& 1.1\times 10^{-5}
\left( \frac{100}{m_{\tilde{\ell}}(\mbox{GeV})}\right)^2.
\ee
This result shows that the BRs of dangerous 
lepton flavor violating  processes like 
$\mu \to e + \gamma$ do not exceed the stringent bounds
as long as $m_{\tilde{\ell}}>170{\mbox{GeV}}$.
As results of the above analyses, it may be said that 
the lepton flavor violation as well as the CP violation 
can be less than their present experimental bounds, in the
case of $U_{eL}$ containing only small mixings.

\subsection{The squark sector}
As in the leptonic sector (\ref{scalarmass}), the squark 
soft mass matrices turn out at $M_{\rm SUSY}$ to be
\be  
\frac{{\bf \tilde{m}^2}_{QLL}}{M^2}& =  &
%m^2_{\tilde{\ell}}
\left( 
\begin{array}{ccc}
10.542 \pm 0.002&0\pm 0.002(1+I) & 0\pm 0.001(1+I)  \\
0\pm 0.002(1+I)  & 10.542 \pm 0.002
& 0\pm 0.001(1+I)  \\
 0\pm 0.001(1+I)  & 0\pm 0.001(1+I)
 &8.192 \pm 0.002 
\end{array}
\right),\nn \\
\frac{{\bf \tilde{m}^2}_{uRR}}{M^2} &=&
%m^2_{\tilde{\ell}}
\left( 
\begin{array}{ccc}
10.092 \pm 0.002 &0\pm 0.002(1+I)
& 0\pm 0.001(1+I)  \\
0\pm 0.002(1+I)  & 10.092 \pm 0.002
& 0\pm 0.001(1+I)\\
0\pm 0.001(1+I) & 0\pm 0.001(1+I)
&5.7878\pm 0.002  
\end{array}
\right),\nn \\ 
\frac{{\bf \tilde{m}^2}_{dRR}}{M^2} &=&
%m^2_{\tilde{\ell}}
\left( 
\begin{array}{ccc}
9.741 \pm 0.002 &0\pm 0.002(1+I)
& 0\pm 0.001(1+I)  \\
0\pm 0.002(1+I)  & 9.741 \pm 0.002
& 0\pm 0.001(1+I)\\
0\pm 0.001(1+I) & 0\pm 0.001(1+I)
&9.678\pm 0.001  
\end{array}
\right).
\label{squarkmass} 
\ee  
We can obtain $\delta$'s by estimating
$U_{uL(uR)}^{\dag}{\bf \tilde{m}^2}_{QLL(uRR)} U_{uL(uR)}$ and
$U_{dL(dR)}^{\dag}{\bf \tilde{m}^2}_{QLL(dRR)} U_{dL(dR)}$.
Also here we assume that
\be
U_{uR} = U_{uL} =U_{dL} =V_{\rm CKM}~,~U_{dR} = V_{\rm MNS}^{\dag},
\ee
therefore $(\delta^u)_{LL}=(\delta^d)_{LL}$.
%where the mixing matrix $V_{\rm CKM}$ is given by
%\be
%V_{\rm CKM} &=&\left(
%\begin{array}{ccc}
%0.98  &  0.22 & 0.003 \\
%-0.22  & 0.97  & 0.04 \\
%0.01  & -0.04  & 1
%\\
%\end{array}\right), \\
%\label{UuL}
%\ee

Taking into account the FCNC contribution from both (i) and (ii),
we find that
\be
\mbox{Re}[(\delta^{u,d}_{12})_{LL}]
&\sim & 2.6 \times 10^{-4}~,~
\mbox{Im}[(\delta^{u,d}_{12})_{LL}]
\sim 1.6 \times 10^{-4}~,\\
%\left( \frac{500}{m_{\tilde{q}}(\mbox{GeV})}\right)^2, \\
\mbox{Re}[(\delta^{u,d}_{13})_{LL}]
&\sim & 1.3 \times 10^{-3}~,~
\mbox{Im}[(\delta^{u,d}_{13})_{LL}]
\sim 1.2 \times 10^{-4}~,\\
%\left( \frac{500}{m_{\tilde{q}}(\mbox{GeV})}\right)^2, \\
\mbox{Re}[(\delta^{u,d}_{23})_{LL}]
&\sim & 1.1 \times 10^{-2}~,~
\mbox{Im}[(\delta^{u,d}_{23})_{LL}]
\sim 1.2 \times 10^{-4}~,\\
%\left( \frac{500}{m_{\tilde{q}}(\mbox{GeV})}\right)^2.
%\mbox{Re}\left[(\delta^d_{12})_{LL}\right]
%&\sim & 2.6 \times 10^{-4}~,~
%\mbox{Im}\left[(\delta^d_{12})_{LL}\right]
%\sim 1.6 \times 10^{-4}~,\\
%\left( \frac{500}{m_{\tilde{q}}(\mbox{GeV})}\right)^2, \\
%\mbox{Re}\left[(\delta^d_{13})_{LL}\right]
%&\sim &  1.3\times 10^{-3}~,~
%\mbox{Im}\left[(\delta^d_{13})_{LL}\right]
%\sim 1.2\times 10^{-4}~,\\
%\left( \frac{500}{m_{\tilde{q}}(\mbox{GeV})}\right)^2, \\
%\mbox{Re}\left[(\delta^d_{23})_{LL}\right]
%&\sim &  1.1 \times 10^{-2}~,~
%\mbox{Im}\left[(\delta^d_{23})_{LL}\right]
%\sim 1.2 \times 10^{-4}~,\\
%\left( \frac{500}{m_{\tilde{q}}(\mbox{GeV})}\right)^2.
\mbox{Re}\left[(\delta^u_{12})_{RR}\right]
&\sim &  3.5 \times 10^{-4}~,~
\mbox{Im}\left[(\delta^u_{12})_{RR}\right]
\sim 1.8 \times 10^{-4}~,\\
%\left( \frac{500}{m_{\tilde{q}}(\mbox{GeV})}\right)^2, \\
\mbox{Re}\left[(\delta^u_{13})_{RR}\right]
&\sim & 2.6 \times 10^{-3}~,~
\mbox{Im}\left[(\delta^u_{13})_{RR}\right]
\sim 1.2 \times 10^{-4} ~,\\
%\left( \frac{500}{m_{\tilde{q}}(\mbox{GeV})}\right)^2, \\
\mbox{Re}\left[(\delta^u_{23})_{RR}\right]
&\sim & 2.2 \times 10^{-2}~,~
\mbox{Im}\left[(\delta^u_{23})_{RR}\right]
\sim 1.1 \times 10^{-4}~,\\
%\left( \frac{500}{m_{\tilde{q}}(\mbox{GeV})}\right)^2.
\mbox{Re}\left[(\delta^d_{12})_{RR}\right]
&\sim &  1.9 \times 10^{-3}~,~
\mbox{Im}\left[(\delta^d_{12})_{RR}\right]
\sim 2.2 \times 10^{-4} ~,\\
%\left( \frac{500}{m_{\tilde{q}}(\mbox{GeV})}\right)^2, \\
\mbox{Re}\left[(\delta^d_{13})_{RR}\right]
&\sim &  2.6 \times 10^{-3} ~,~
\mbox{Im}\left[(\delta^d_{13})_{RR}\right]
\sim 2.6\times 10^{-4}~,\\
%\left( \frac{500}{m_{\tilde{q}}(\mbox{GeV})}\right)^2, \\
\mbox{Re}\left[(\delta^d_{23})_{RR}\right]
&\sim & 2.6 \times 10^{-3}~,~
\mbox{Im}\left[(\delta^d_{23})_{RR}\right]
\sim 2.6 \times 10^{-4}~.
%\left( \frac{500}{m_{\tilde{q}}(\mbox{GeV})}\right)^2 .
\ee
%where the A-parameters $A_i$ are assumed to be $O(500\mbox{GeV})$.
The upper bounds for $\delta$'s coming  from the measurements of
$K-\bar{K}$, $D-\bar{D}$, $B_d-\bar{B}_d$  mixing,
$\epsilon_K$, $b \rightarrow s~ \gamma$ and
$\epsilon'/\epsilon$ \cite{fcnc}
are shown in TABLE IV, where
the imaginary parts are constrained by CP-violating processes.
It is seen that
$\delta$'s given above satisfy well the experimental constraints 
except for
$\sqrt{|\mbox{Im}(\delta^d_{12})_{LL}(\delta^d_{12})_{RR}|}$,
 which is comparable to the constraint.

\begin{table}[htb]
\begin{center}
\caption{Limits on the $|\delta^{d(u)}_{ij}|$ from
$K-\bar{K}$, $D-\bar{D}$, $B_d-\bar{B}_d$  mixing,
$\epsilon_K$, $b \rightarrow s~ \gamma$ and
$\epsilon'/\epsilon$ for
$m_{\tilde{g}}^2/m_{\tilde{q}}^2=1$ \cite{fcnc}.
Here the parameter $\tilde{m}_{\tilde{q}}$ denotes
$m_{\tilde{q}}(\mbox{GeV})/500$.}
\begin{tabular}{|c|c|c|} \hline \hline
$\sqrt{|\mbox{Re}(\delta^d_{12})^2_{LL,RR}|}$
& $\sqrt{|\mbox{Re}(\delta^d_{12})_{LL}(\delta^d_{12})_{RR}|}$
& $\sqrt{|\mbox{Re}(\delta^d_{12})^2_{LR}|}$
\\ \hline
$4.0 \times 10^{-2} ~\tilde{m}_{\tilde{q}}$
& $2.8 \times 10^{-3} ~\tilde{m}_{\tilde{q}}$
& $4.4 \times 10^{-3} ~\tilde{m}_{\tilde{q}}$
\\ \hline \hline
$\sqrt{|\mbox{Re}(\delta^d_{13})^2_{LL,RR}|}$
& $\sqrt{|\mbox{Re}(\delta^d_{13})_{LL}(\delta^d_{13})_{RR}|}$
& $\sqrt{|\mbox{Re}(\delta^d_{13})^2_{LR}|}$
\\ \hline
$9.8 \times 10^{-2} ~\tilde{m}_{\tilde{q}}$
& $1.8 \times 10^{-2} ~\tilde{m}_{\tilde{q}}$
& $3.3 \times 10^{-3} ~\tilde{m}_{\tilde{q}}$
\\ \hline \hline
$\sqrt{|\mbox{Re}(\delta^u_{12})^2_{LL,RR}|}$
& $\sqrt{|\mbox{Re}(\delta^u_{12})_{LL}(\delta^u_{12})_{RR}|}$
& $\sqrt{|\mbox{Re}(\delta^u_{12})^2_{LR}|}$
\\ \hline
$1.0 \times 10^{-1} ~\tilde{m}_{\tilde{q}}$
& $1.7 \times 10^{-2} ~\tilde{m}_{\tilde{q}}$
& $3.1 \times 10^{-3} ~\tilde{m}_{\tilde{q}}$
\\ \hline \hline
$\sqrt{|\mbox{Im}(\delta^d_{12})^2_{LL,RR}|}$
& $\sqrt{|\mbox{Im}(\delta^d_{12})_{LL}(\delta^d_{12})_{RR}|}$
& $\sqrt{|\mbox{Im}(\delta^d_{12})^2_{LR}|}$
\\ \hline
$3.2 \times 10^{-3}~ \tilde{m}_{\tilde{q}}$
& $2.2 \times 10^{-4} ~\tilde{m}_{\tilde{q}}$
& $3.5 \times 10^{-4} ~\tilde{m}_{\tilde{q}}$
\\ \hline \hline
$|(\delta^d_{23})_{LL,RR}|$
& 
& $|(\delta^d_{23})_{LR}|$
\\ \hline
$8.2~ \tilde{m}_{\tilde{q}}^2$
&
& $1.6 \times 10^{-2} ~\tilde{m}_{\tilde{q}}^2$
\\ \hline \hline
$|\mbox{Im}(\delta^d_{12})_{LL,RR}|$
& 
& $|\mbox{Im}(\delta^d_{12})_{LR}|$
\\ \hline
$4.8 \times 10^{-1} ~\tilde{m}_{\tilde{q}}^2$
& 
& $2.0 \times 10^{-5} ~\tilde{m}_{\tilde{q}}^2$
\\ \hline \hline
\end{tabular}
\end{center}
\end{table}

The mixing masses between the left-handed and 
right-handed squarks
and also their effects to FCNC and CP-violation may be
evaluated just as done for the slepton sector.
The A-parameters at the weak scale are given by
\be 
\frac{a_u}{M}&=&
\left( 
\begin{array}{ccc}
a & a & b \\
a & a & b \\
b & b & c 
\end{array}\right),~~~
\begin{array}{l}
a=(4.929 \pm 0.020)+I~(0 \pm 0.017) \\
b=(2.701 \pm 0.007)+I~(0 \pm 0.005) \\
c=(2.705 \pm 0.001)+I~(0 \pm 0.002),\\
\end{array}
\label{au}\\ 
\frac{a_d}{M}&=&
\left( 
\begin{array}{ccc}
a & a & b \\
a & a & b \\
b & b & c 
\end{array}\right),~~~
\begin{array}{l}
a=(6.241 \pm 0.020)+I~(0 \pm 0.017) \\
b=c=(5.391 \pm 0.013)+I~(0 \pm 0.016),
\end{array}
\label{ad}
\ee
The indices $\delta$, which should be compared with the 
experimental constraints shown also in TABLE IV, 
are found to be
\be
\mbox{Re}\left[(\delta^u_{12})_{LR}\right]
&\sim & 3.9 \times 10^{-4}~\tilde{m}_{\tilde{q_u}}^{-2} ~,~
\mbox{Im}\left[(\delta^u_{12})_{LR}\right]
\sim 4.1 \times 10^{-5}~\tilde{m}_{\tilde{q}_u}^{-2} ~,\\
%\left( \frac{500}{m_{\tilde{q}}(\mbox{GeV})}\right)^2, \\
\mbox{Re}\left[(\delta^u_{13})_{LR}\right]
&\sim & 7.7 \times 10^{-5}~\tilde{m}_{\tilde{q}_u}^{-2} ~,~
\mbox{Im}\left[(\delta^u_{13})_{LR}\right]
\sim 6.2 \times 10^{-5}~\tilde{m}_{\tilde{q}_u}^{-2} ~,\\
%\left( \frac{500}{m_{\tilde{q}}(\mbox{GeV})}\right)^2, \\
\mbox{Re}\left[(\delta^u_{23})_{LR}\right]
&\sim & 6.5 \times 10^{-4}~\tilde{m}_{\tilde{q}_u}^{-2}~,~
\mbox{Im}\left[(\delta^u_{23})_{LR}\right]
\sim 2.3 \times 10^{-4}~\tilde{m}_{\tilde{q}_u}^{-2}~,\\
%\left( \frac{500}{m_{\tilde{q}}(\mbox{GeV})}\right)^2.
\mbox{Re}\left[(\delta^d_{12})_{LR}\right]
&\sim & 3.9 \times 10^{-5}~\tilde{m}_{\tilde{q}_d}^{-2} ~,~
\mbox{Im}\left[(\delta^d_{12})_{LR}\right]
\sim 1.4 \times 10^{-5}~\tilde{m}_{\tilde{q}_d}^{-2} ~,\\
%\left( \frac{500}{m_{\tilde{q}}(\mbox{GeV})}\right)^2, \\
\mbox{Re}\left[(\delta^d_{13})_{LR}\right]
&\sim &  1.5 \times 10^{-4}~\tilde{m}_{\tilde{q}_d}^{-2} ~,~
\mbox{Im}\left[(\delta^d_{13})_{LR}\right]
\sim 1.8 \times 10^{-4} ~\tilde{m}_{\tilde{q}_d}^{-2}~,\\
%\left( \frac{500}{m_{\tilde{q}}(\mbox{GeV})}\right)^2, \\
\mbox{Re}\left[(\delta^d_{23})_{LR}\right]
&\sim &  2.2 \times 10^{-4}~\tilde{m}_{\tilde{q}_d}^{-2}~,~
\mbox{Im}\left[(\delta^d_{23})_{LR}\right]
\sim 2.6 \times 10^{-4}~\tilde{m}_{\tilde{q}_d}^{-2}~,
\ee
where $\tilde{m}_{\tilde{q}_{u(d)}}^{-2}=
(500/m_{\tilde{q}_{u(d)}}(\mbox{GeV}))^2$.
For squarks, all of them are small enough to suppress 
the FCNC and the CP violation within the bounds,
although we have assumed $U_{dR}$ to be the bi-maximal 
mixing matrix.

\section{Conclusion}
In this paper, we investigated how much the infra-red attractive force
of gauge interactions can soften  the SUSY flavor problem
in the orbifold GUT of Kawamura \cite{kawamura1}.
First we discussed the notion of gauge coupling unification
in the orbifold GUT models, where the unified gauge symmetry
is explicitly broken by the boundary conditions.
It is natural for the bulk theory to recover the unified symmetry
as the scale goes much shorter than the radius of compactified
dimensions. 
We showed explicitly that the running gauge couplings defined
in the extra-dimensional sense approach to each other 
asymptotically (asymptotic unification).
In the four-dimensional picture, this occurs due to the power-law
running behavior of the gauge couplings.

The radiative corrections by the bulk gauge fields make
the SSB parameters subject to the power-law running also.
Then the ratio of the SSB parameters to the
gaugino mass at the compactification scale are fixed to
their infra-red attractive fixed point values,
which are totally flavor universal \cite{kubo6}.
It should be noted that the SSB parameters at low energy
are also fixed solely by the gaugino mass scale,
and, therefore, insensitive to those in the fundamental theory.
Thus this suggests an interesting possibility 
for the SUSY flavor problem and the CP problem.

We examined the one-loop RG flows for the general soft SSB
parameters in the orbifold GUT models.
Then the Yukawa couplings to the bulk
Higgs fields, which also show power-law running behavior,
split the SSB parameters of the first two and the third 
generations.
In the calculations we neglected the logarithmic corrections
including the breaking effects of the GUT symmetry due to 
boundary conditions, since the most dominant flavor dependence
comes from the corrections due to the bulk Yukawa couplings.
We assumed also $M_{\rm PL}/M_c \sim 10^3$.

Now there are two sources for the flavor violating masses of the 
SUSY particles; (i) flavor dependence in the fixed points
induced by the Yukawa couplings, (ii) deviation from the fixed
point values due to finite radius of the compactified dimensions.
As for (ii), we found that the arbitrary disorder in the SSB 
parameters at the fundamental scale are sufficiently suppressed 
at $M_{\rm GUT}$.
Therefore this effect does not cause any problems in FCNC 
processes or in dangerous CP-violating phenomena, since
the fixed points are real.
So what we should be concerned more is the effect (i)
in the orbifold GUT models.

The key ingredients for the FCNC processes are the flavor changing
elements of the mass matrices of squarks and sleptons obtained 
after rotation to the basis of mass eigenstates for quarks and 
leptons. However the rotation matrices are unknown, though the
mixing matrices $V_{\rm CKM}$ and $V_{\rm NMS}$ are given 
experimentally. We first assume that the rotation matrices of
the fields belonging to ${\bf 10}$ of SU(5) GUT are given by 
$V_{\rm CKM}$, while the rotation matrices of the fields belonging 
to $\overline{\bf 5}$ are $V_{\rm NMS}$ (\ref{eReL}). 
This would be regarded as the maximal estimation of the rotations.
Then it is found that the indices of the off-diagonal elements of
the soft scalar masses $\delta_{LL}$ and $\delta_{RR}$ are both
suppressed sufficiently. The reason of this is as follows.
Indeed splitting of the fixed points values for the third 
generation to others are sizable for the fields in ${\bf 10}$
due to large $Y_t$. However the rotation matrix $V_{\rm CKM}$ contains
only small mixings.
On the other hand the degeneracy of the fixed points for the fields
in $\overline{\bf 5}$ is fairly good. Therefore the off-diagonal
elements remain tiny, even if the mass matrices are transformed by
the large mixing matrix $V_{\rm MNS}$. 

Unfortunately, however, it is found to be hard for the A-parameters to
satisfy the experimental constraints under the above assumption
on  the rotation matrices. 
The index $(\delta^{\ell}_{12})_{LR}$ in the slepton sector
appears exceeding the present experimental bound, though
all the other indices in the both sectors are lower than their
limit.
However, if the rotation matrix $U_{eL}$ is a small mixing
one like $V_{\rm CKM}$, (namely $V_{\rm MNS} \sim U_{\nu}$),
then the index $(\delta^{\ell}_{12})_{LR}$ is found to 
become lower than the constraint unless the slepton masses
are very light.
Thus our mechanism can soften the SUSY flavor problem and also
the CP problem.
After all we conclude that
the  mechanism of \cite{kubo6} to solve the
SUSY flavor problem may be combined with
the  mechanism of \cite{kawamura1} to overcome
the doublet-triplet splitting problem in extra dimensions.

In this paper we have not discussed the case with the 
right-handed neutrino. It has been well-known that the
Yukawa coupling between the lepton-doublet and the right-handed
neutrino generates sizable mixings in the slepton masses
in comparison with the current bounds for the lepton
violating processes \cite{bm,LFV}, unless the neutrino Yukawa
coupling is rather small.
This effect is caused by the large mixing angles of
$V_{\rm MNS}$. In the orbifold GUT, the fixed points are not
degenerate, therefore the off-diagonal elements would be 
generated more due to the large mixings.
Also it should be concerned also that the running above the 
GUT scale is affected by the neutrino Yukawa coupling \cite{GUTmixing}.
Then larger mass mixings could appear in $\bar{\bf 5}$-sector, therefore 
not only the sleptons but also the squarks sectors should be reanalyzed.
Here we would like to leave these problems to future studies.

\acknowledgments

This work is supported by the Grants-in-Aid
for Scientific Research  from
the Japan Society for the Promotion of Science (JSPS) (No. 11640266,
No. 13135210, No. 13640272).
We would like to thank
H.~Nakano and T.~Kobayashi
for useful discussions.

%\begin{references}

%\end{references}

\end{document}